\documentclass[pre,twocolumn,superscriptaddress,aps,notitlepage,10pt]{revtex4-1}
\usepackage{natbib}
\usepackage{times}
\usepackage{amssymb}
\usepackage{amsmath}
\usepackage{bm}
\usepackage{color}
\usepackage[colorlinks=true,urlcolor=blue,breaklinks,citecolor=blue]{hyperref}
\usepackage{graphicx}
\usepackage{braket}
\usepackage{subfigure}
\usepackage{xcolor}
\usepackage{soul}

\usepackage[utf8]{inputenc}
\usepackage{hyperref}
\definecolor{fuchsia}{rgb}{1.0, 0.0, 1.0}

\newcommand{\kat}{\textcolor{blue}}

\begin{document}

\title{The effect of the processing and measurement operators on the expressive power of quantum models}

\author{Aikaterini (Katerina) Gratsea}
\affiliation{ICFO-Institut  de  Ciencies  Fotoniques,  The  Barcelona  Institute  of Science  and  Technology, 08860  Castelldefels  (Barcelona),  Spain}
\email[]{gratsea.katerina@gmail.com}
\author{Patrick Huembeli}
\affiliation{Menten AI, 1160 Battery Street East, Suite 100
San Francisco, CA 94111, USA}

\begin{abstract}
 \textbf{Abstract} There is an increasing interest in Quantum Machine Learning (QML) models, how they work and for which applications they could be useful. There have been many different proposals on how classical data can be encoded and what circuit ans\"atze and measurement operators should be used to process the encoded data and measure the output state of an ansatz. The choice of the aforementioned operators plays a determinant role in the expressive power of the QML model. In this work we investigate how certain changes in the circuit structure change this expressivity. We introduce both numerical and analytical tools to explore the effect that these operators have in the overall performance of the QML model. These tools are based on previous work on the teacher-student scheme, the partial Fourier series and the averaged operator size. We focus our analysis on simple QML models with two and three qubits and observe that increasing the number of parameterized and entangling gates leads to a more expressive model for certain circuit structures. Also, on which qubit the measurement is performed affects the type of functions that QML models could learn. This work sketches the determinant role that the processing and measurement operators have on the expressive power of simple quantum circuits.
\end{abstract}
\maketitle

\section{Introduction}
In recent years, there is increasing interest in the capabilities of Quantum Machine Learning (QML) models and their potential applications. To better understand for what tasks a QML model could be potentially used for~\cite{modern_applications}, we first need to understand where its strengths and weaknesses 
lie~\cite{goal_of_QML, Alejandro_new, Maria_kernel, Sim_2019, paper_storage, TS_scheme, data_re-uploading, Lewenstein_2021, Sim_2019, new1, NEW2}. A first step to better understand what a QML model is capable of, is to study its expressivity. It has been shown that a specific quantum circuit architecture can be used as a universal function approximator~\cite{data_re-uploading}, i.e. it can approximate any classification function up to arbitrary precision. The work of~\cite{schuld2020effect} further sheds light on how the encoding of the data fundamentally limits the expressivity of quantum models. It also verifies that parameterized QML models with arbitrary single rotations and entangling gates give rise to complex trigonometric functions~\cite{schuld2020effect, huang2020power}.

Instead of focusing on the data encoding, other works study the importance of the trainable part of the quantum circuits (consisting of data independent but parameterized gates) to which we refer to as the processing part of the circuit. For example, in~\cite{Ostaszewski_2021}, the authors perform a structure optimization that aims to find a more hardware efficient architecture for the desired problem. Finally, recent works focus on benchmarking different circuit structures both numerically~\cite{ancilla_best, Sim_2019, qVol_num, TS_scheme} and theoretically~\cite{scrambling, Abbas_2021, qVol_theo, paper_storage}.

How to choose an ansatz for a supervised QML task is still an open question. Circuit ans\"atze that are used to find ground states in quantum chemistry~\cite{Cao_2019} or many-body physics are often inspired by the problem itself~\cite{nature, symmetry, Cai_2020} and some understanding of the optimization problem at hand can help to find a well-suited ansatz~\cite{Ostaszewski_2021, Economou2}. For QML applications this does not apply in general~\cite{VQE1, PhysRevA.98.032309, PhysRevA.101.032308, Economou} which might partially be because we still do not fully understand how specific parts of the processing and measurement operators affect the performance of the QML model. Moreover, even for small quantum circuits with a few number of qubits, there are still some open questions: "How many parameterized gates should be used?", "How many entangling gates and where?", "Where should the circuit be measured?" and "Do ancilla qubits help improve the expressivity of the circuit?".

Inspired by these questions, we aim to study the effect of the processing part of QML architectures. We believe that the structure of the processing and measurement operators play a determinant role in the QML model performance. Here the term performance characterizes the model's ability to learn through training~\cite{qBas}, the type of functions that it can express~\cite{schuld2020effect} and its scrambling capability~\cite{scrambling, info_scrambling}, i.e. extracting information from the input state to the readout qubit.

We use both analytical and numerical tools to explore how the processing unitaries and measurement affect the properties of a QML model. We focus our analysis on simple quantum models with just 2-3 qubits, which help us better understand how different parts of the processing and measurement operators affect the performance. This analysis allows us to understand how the processing architecture affects the expressivity of the whole model, at least for circuits with small amounts of qubits, with the hope to draw insights on how specific parts will affect the performance of more complicated architectures with more qubits. We observe that increasing the number of parameterized and entangling gates leads to a more expressive model for certain circuit structures and that ancillary qubits help increasing the expressive power of the model. Finally, applying the measurement on certain qubits changes the classes of functions that QML models can learn.

The paper is structured as follows. In section~\ref{Theotools}, we introduce the analytical and numerical tools discussed in this work. In section~\ref{analysis}, we apply these tools to analyze the performance of two simple quantum models. In section~\ref{Fourier section}, we discuss that regarding the processing and measurement operator as a single unit could give insights on the performance of the QML models. Finally, in Section~\ref{conclusions}, we summarize the importance of the non-encoding unitaries and the effect they have on the expressive power of quantum models. 
\section{Tools for analysing QML models}\label{Theotools}
\begin{figure}[h]
\includegraphics[width=\columnwidth]{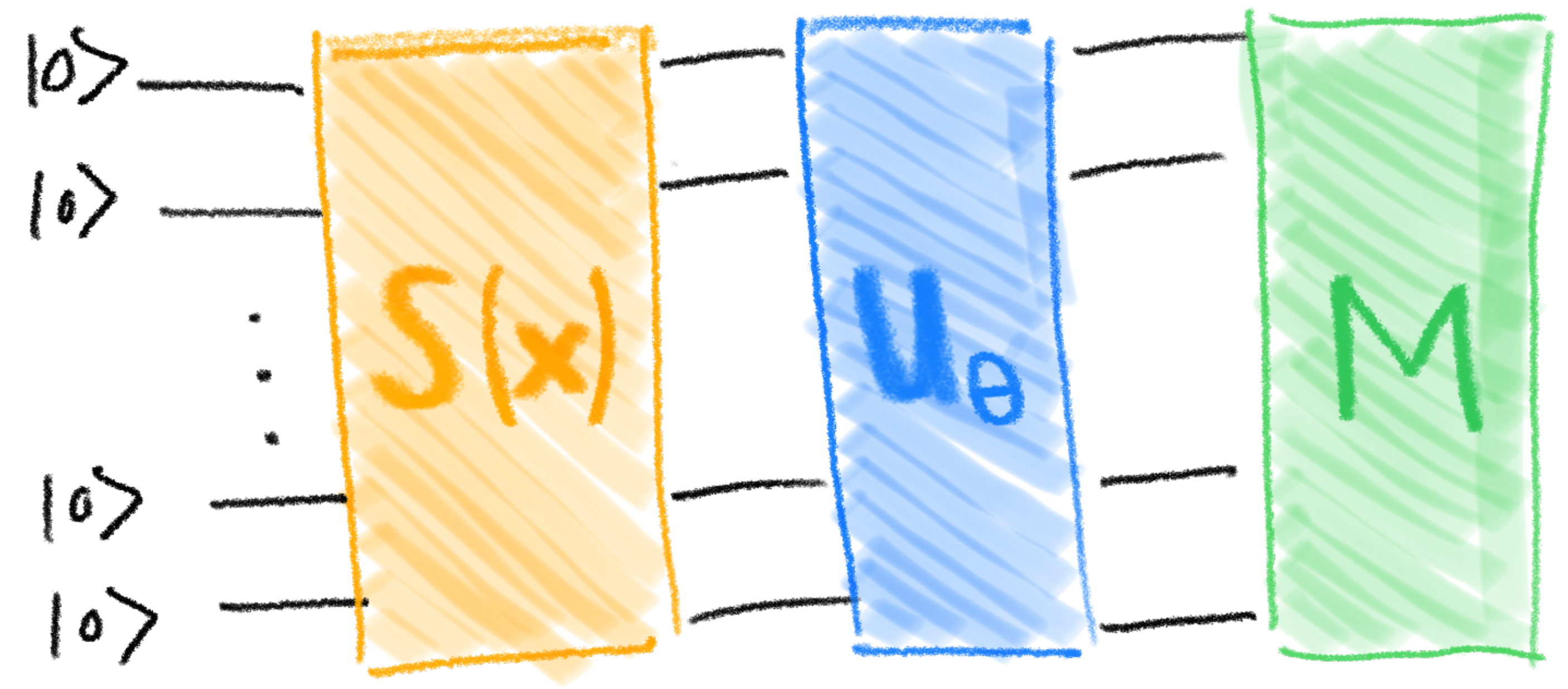}
\caption{The building blocks for an arbitrary quantum circuit: encoding (yellow unitary), processing (blue parameterized unitary) and measurement (green operator).}
\label{parts} 
\end{figure}

We study quantum circuits that consist of three parts: the encoding unitaries $S \left( x \right)$, the processing gates $U_{\theta}$ (consisting of data independent
but parameterized gates) and the measurement operator $M$. The building blocks of such arbitrary quantum circuits are shown in Fig.~\ref{parts}. 
Throughout this work, we focus on supervised learning tasks with a given data set $\mathcal{D} = \{(\bm{x}^k, y^k) \} $, where the input data $\bm{x}^k$ are encoded by $S \left( \bm{x}^k \right)$ and $y^k$ are the labels given by the measurement outcome. The initial state is $\ket{0}^{\otimes N}$, where $N$ is the number of qubits. As shown in Fig.\ref{parts}, we first apply the encoding unitary $S \left( \bm{x} \right)$ (yellow), then the processing $U_{\theta}$ (blue) and finally the measurement operator $M$ (green). 

In this section, we focus on two theoretical tools for characterizing the performance of QML models, i.e. the averaged operator size~\cite{scrambling} and the Fourier representation~\cite{schuld2020effect}. These methods give us quantitative results on the effect of the processing and measurement operators (blue parameterized unitary and green operator in Fig.\ref{parts}). We also employ an extended version of the teacher-student scheme discussed in~\cite{TS_scheme} to quantify differences in the prediction maps numerically, i.e. the density plots of the model predictions or labels $y^k$ for the input data $\bm{x}^k$.
\subsection{Averaged operator size}\label{av_theory}
In the works~\cite{scrambling, Nahum_2018, Roberts_2018}, the authors introduce related quantities to characterize quantum information scrambling, i.e.  extracting information from the input state $S(\bm{x}) |\bm{0} \rangle$ to the readout qubit. In the work of~\cite{scrambling}, the authors propose the averaged operator size - a quantity that depends on the circuit architecture, the processing and the measurement operators. Moreover, its value is positively correlated with the learning efficiency of the QNN architecture.

To calculate the averaged operator size of an arbitrary operator $\hat{O}$, we decompose it into a summation of a Pauli strings, which can be done for any Hermitian operator~\cite{NC_book}. Such a decomposition has the form
\begin{equation}\label{Op_in_paulis}
\hat{O}=\sum_{\boldsymbol{\alpha}} c_{\boldsymbol{\alpha}} \hat{\sigma}_{\alpha_{1}}^{1} \otimes \hat{\sigma}_{\alpha_{2}}^{2} \cdots \otimes \hat{\sigma}_{\alpha_{n}}^{n},
\end{equation}
where $\hat{\sigma}^i_{\alpha_i}$ for $\alpha_i \in\{0,1,2,3\}$ define the Pauli operators including the identity acting on qubit i. The coefficients $c_\alpha$ can be computed by
\begin{equation}
c_\alpha=\frac{1}{2^{n}} \operatorname{Tr}\left(O  \hat{\sigma}_{\alpha_{1}}^{1} \otimes \hat{\sigma}_{\alpha_{2}}^{2} \cdots \otimes \hat{\sigma}_{\alpha_{n}}^{n}\right).
\end{equation}
The operator size of a Hermitian operator $\hat{O}$ is given by
\begin{equation}
\operatorname{Size}(\hat{O})=\sum_{\boldsymbol{\alpha}}\left|c_{\boldsymbol{\alpha}}\right|^{2} l(\boldsymbol{\alpha}),
\end{equation}
where $c_{\boldsymbol{\alpha}}$ are the coefficients of decomposition in Eq.~\eqref{Op_in_paulis} and $l(\boldsymbol{\alpha})$ counts the number of non-identity matrices in each Kronecker product in the summation given by the same equation. In our case, the hermitian operator $\hat{O} = \hat{{U_{\theta}}}^{\dagger} \hat{M} \hat{\mathcal{U_{\theta}}}$ depends only on the processing unitary $U_{\theta}$ and the measurement $M$ operators. 

To obtain the averaged operator size for a given circuit architecture $U_{\theta}$ and a fixed measurement $M$, we take the average over Haar random unitaries $U_{\theta}$
\begin{equation} \label{overlina_Size}
\overline{\operatorname{Size}}=\int d \hat{{U_{\theta}}} \operatorname{Size}\left(\hat{{U_{\theta}}}^{\dagger} \hat{M} \hat{U_{\theta}}\right).
\end{equation}
A larger value of the average operator size suggests a more expressive circuit structure~\cite{scrambling}.

\subsection{Map differencies from teacher-student scheme}\label{TS_theory}
The teacher student scheme employed in~\cite{TS_scheme} can be used to compare the expressivity of different QML circuit architectures. The main idea is that alternately one circuit plays the role of the teacher and the other one of the student. A randomly initialized teacher circuit maps the input data $\{ \bm{x}^k \}$ to labels $y^k$ that are to be learned by the student, and then, the roles are reversed. To compare their performances in the original work~\cite{TS_scheme}, three different quantitative scores were considered: the accuracy score, the (average) loss and the relative entropy and the prediction maps for a more qualitative overview. Here, we add another score to the TS scheme that directly compares differences between prediction maps which allows us to better quantify them. Specifically, we compute the average of the differences between the prediction maps of the teacher and student
\begin{equation}\label{dif}
    \overline{\Delta y} = \dfrac{1}{p} \sum_k^p  | y^k_T  - y^k_S| ,
\end{equation}
where $\{ y^k_T \}$ are the generated labels from the teacher and $\{ y^k_S\}$ the learned labels of the student where $k$ runs over all points $p$ of a given input dataset $\{x^k \}$. Also, $y^k_T, y^k_S$ take continuous values in $[-1, 1]$, but for the calculation of the $\overline{\Delta y}$ we re-scale them to be in $[0, 1]$. We can present the $\overline{\Delta y}$ as a percentage difference between the two studied models. This allows us directly to compare how similar the prediction maps of the two models are. It is a quantitative measure of how well the student can learn the teacher, i.e. a $\overline{\Delta y}$ equal to zero suggests that the student learns the teacher perfectly, while high values for $\overline{\Delta y}$ indicate that the student is not able to learn the labeling provided by the teacher.
\subsection{The representation of quantum models with partial Fourier series}
\label{Fourier}
In the work of~\cite{schuld2020effect}, the authors explore how the data encoding influences the class of functions that a quantum model can learn. Circuit architectures employed in supervised tasks with multiple encoding unitaries on different qubit can be mapped to a partial multivariate Fourier series
\begin{equation}\label{Ffun}
    f_{\boldsymbol{\theta}}(\boldsymbol{x})=\sum_{\boldsymbol{j}} \sum_{\boldsymbol{k}} c_{\boldsymbol{j} \boldsymbol{k}} e^{i \boldsymbol{x} \cdot\left(\boldsymbol{\lambda}_{\boldsymbol{k}}-\boldsymbol{\lambda}_{j}\right)},
\end{equation}
where $j, k \in\left[2^{d}\right]^{N}$ with $N$ the number of qubits and $d$ is the dimension of an encoding gate, i.e. $d=1$ if it is a single qubit gate. As it is explained in the original work, the frequency spectrum determines the functions that the quantum model could express, while the coefficients $c_{\boldsymbol{j} \boldsymbol{k}}$ determine how the accessible functions can be combined. 

The number of Fourier basis functions is solely determined by the eigenvalues of the data-encoding Hamiltonians. This means that repeated data encoding gives rise to a larger frequency spectrum $\Omega$ and more complicated function classes. The processing and the measurement of the circuit determine the coefficients, and therefore, how the accessible functions can be combined. Inspired by the work of~\cite{schuld2020effect}, in Section \ref{Fourier section}, we study how the processing and measurement operators of simple quantum circuits affect the coefficients, and therefore, the final function classes that these quantum circuits "has access to". This analysis sheds light on how individual elements of the processing and measurement architecture affect the function classes that the quantum model can express. 
\begin{figure}
\centering
\includegraphics[width=.4\linewidth]{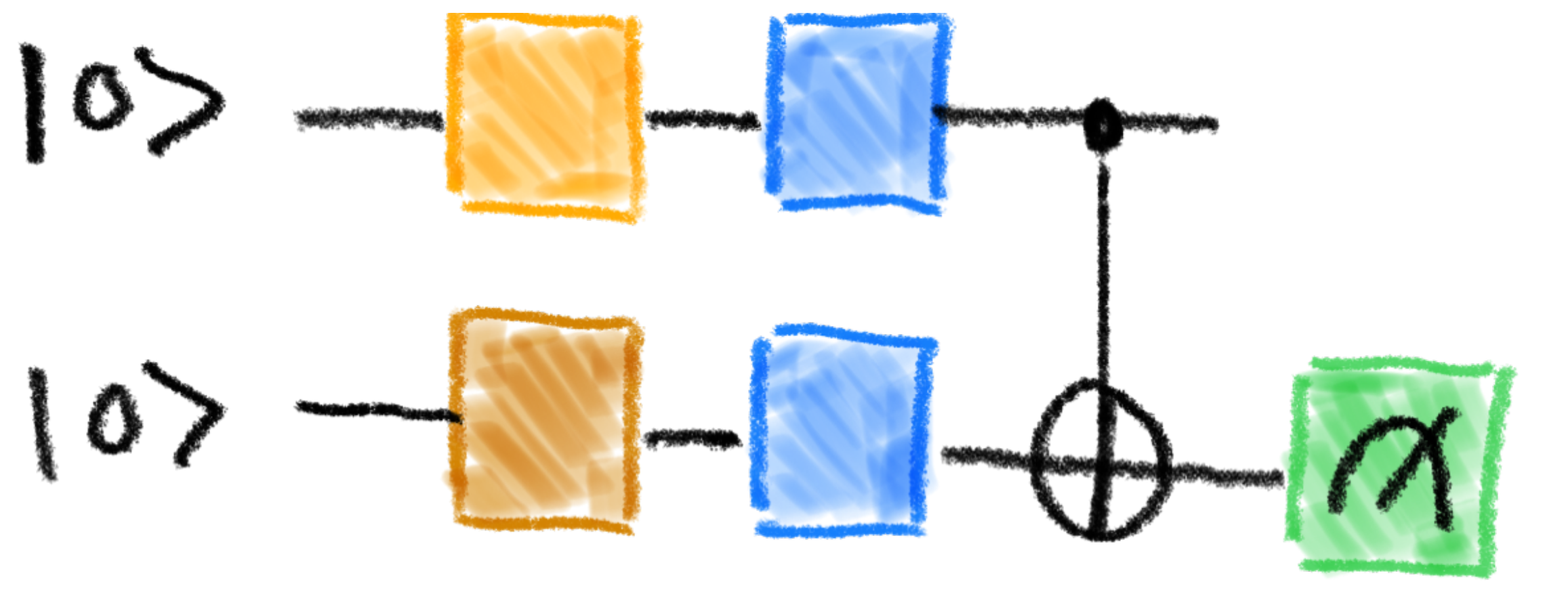}
\includegraphics[width=.4\linewidth]{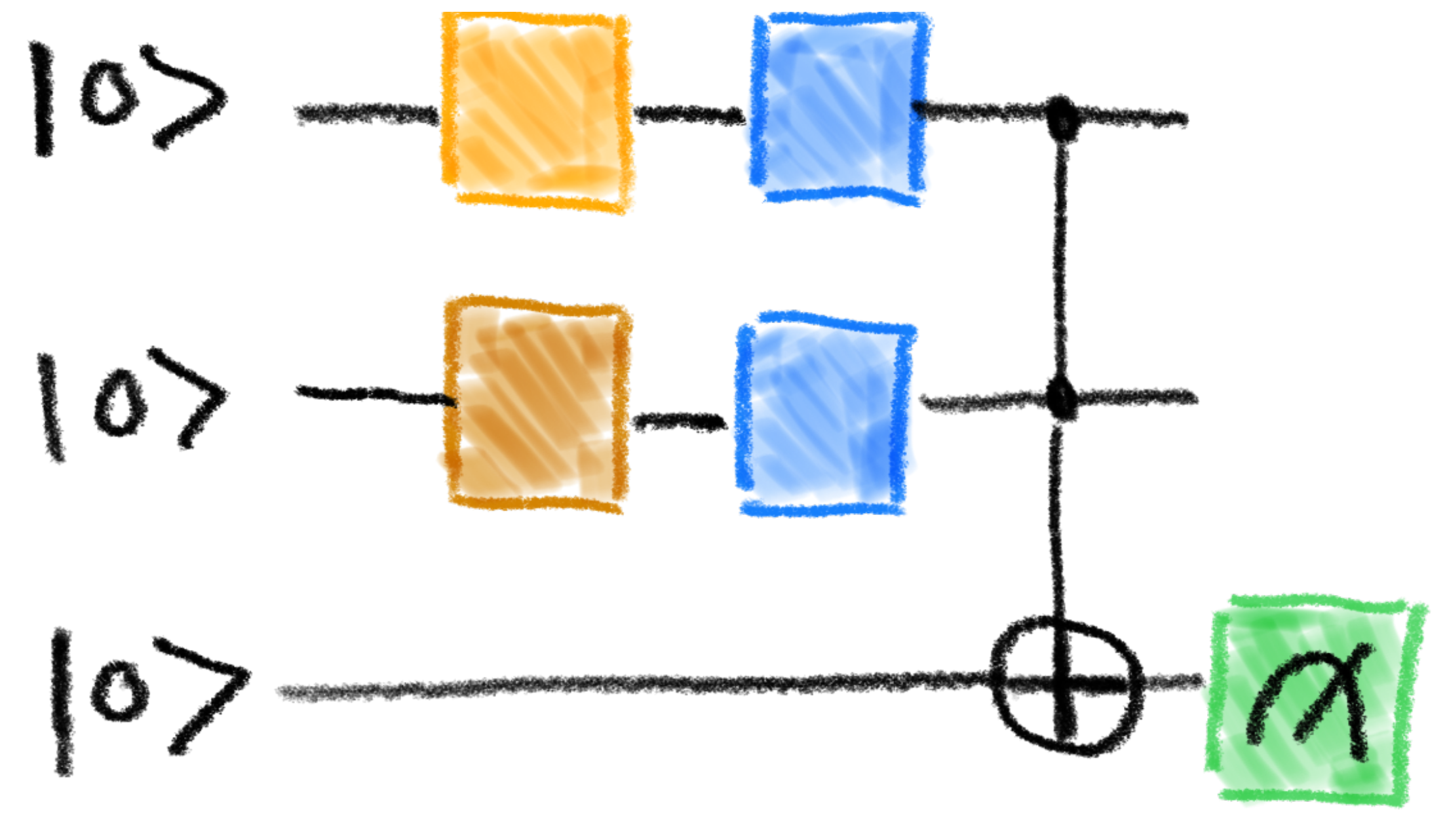}
\caption{A 2-qubit and 3-qubit simple circuit architecture. The light and dark orange colors represent the data encoding gates, while the blue the parameterized gates. A CNOT gate is applied before the measurement operator for the 2-qubit circuit, while a Toffoli gate for the 3-qubit circuit.}
\label{simple_circuits}
\end{figure}
\section{Simple quantum models} \label{analysis}
In this section, we apply the tools described in the Section~\ref{Theotools} on two simple quantum models shown in Fig.~\ref{simple_circuits}. We use a two dimensional input data set with $\bm{x}^k = ( x^k_1, x^k_2 )$ and encode it to the circuit with the gates $S(\bm{x}) = R_x (x_1) \otimes R_x (x_2)$ applied on two distinct qubits, where $R_x(\phi) = \exp( -i\phi \sigma_x/2 )$ is the single qubit $X$ rotation gate. These data encoding gates are depicted in light and dark orange colors respectively in the circuit diagrams. The blue gates represent the parameterized single qubit rotations $ Rot(\phi, \theta, \omega) = R_z(\omega)R_y(\theta)R_z(\phi)$. The predictions take continuous values $y^k \in [-1, 1]$ given by the outcome of the measurement $ \bra{\psi^k} Z \ket{\psi^k} $, where $\ket{\psi^k} = U_{\theta} S(\bm{x}) \ket{0}^{\otimes N}$.

\subsection{Averaged operator size of simple quantum models}\label{av_analysis}
We compute the averaged operator size of Eq.~\eqref{overlina_Size} for the 2-qubit and 3-qubit circuits of Fig.~\ref{simple_circuits} by taking the average over the Haar random unitaries for the parameterized single qubit unitaries depicted with blue color. For the numerical simulation we use a Monte-Carlo integration~\cite{McClean} introduced in the Appendix~\ref{MC}.
In Fig.\ref{av_results}, we plot the averaged operator size as a function of the number of parameterized gates. The error bars are the standard deviation from the mean value. We increase the number of parameterized gates by adding a layer $L$ with two arbitrary single qubit gates (blues color) and an a CNOT entangling gate, $L=CNOT Rot(\phi, \theta, \omega) \otimes Rot(\phi', \theta', \omega')$.  We stress here that the 3-qubit circuit always has a Toffoli gate before the measurement operator. 

\begin{figure}
\includegraphics[width=\columnwidth]{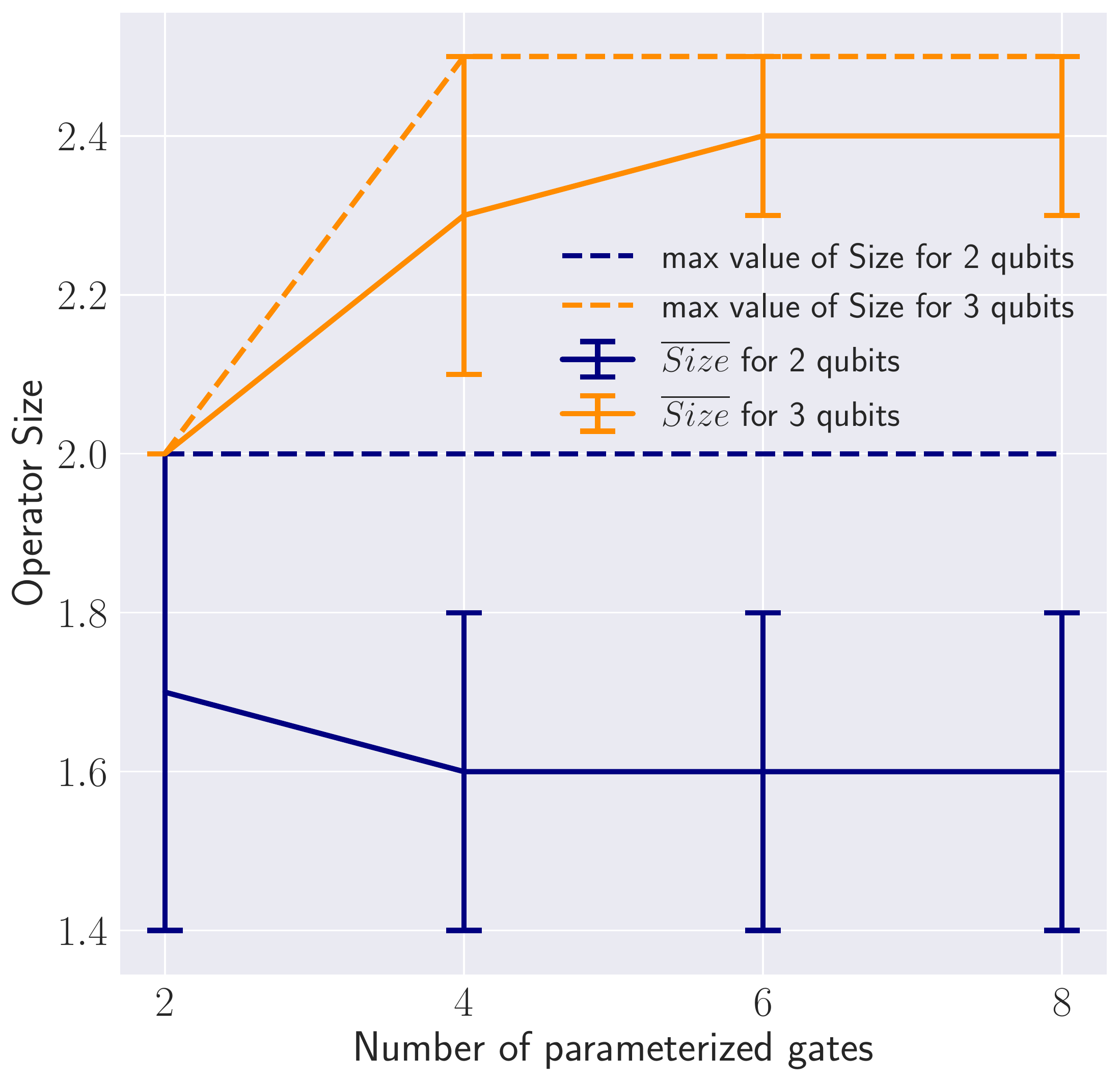}
\caption{We plot the averaged operator sizes defined in Eq.\eqref{overlina_Size} and their standard deviation as we increase the number of parameterized and entangling gates. The orange and blue solid lines are for the MC simulation of 2 and 3 qubits, respectively. The dashed orange and blue solid lines give the maximum value that each operator size could get for a given number of parameterized and entangling gates of 2 and 3 qubits, respectively. These results suggest that the 3-qubit circuit architecture is more expressive than the 2-qubit.}
\label{av_results} 
\end{figure}

The average operator size for the 2-qubit case has approximately the same value around $1.60$ independent of the number of gates used (see blue solid line of Fig.~\ref{av_results}). The drop observed from two to four parameterized gates is small while the error is large. Therefore, this suggests that for the 2-qubit case the performance is approximately the same as we increase the number of parameterized gates.
On the contrary, for the 3-qubit case, the averaged operator size significantly increases once we move from two to four parameterized gates, and afterwards, reaches a plateau. These results already suggest that the 3-qubit case performs better than the 2-qubit case, i.e. the circuit structure of the processing and measurement operator is more expressive.

Next, we focus on the maximum value of the operator size, which corresponds to specific values of the parameterized angles that give the best possible performance of this circuit structure (represented with the dashed line in Fig.~\ref{av_results}). From the dashed lines in Fig.~\ref{av_results}, we see that for both circuits the maximum value of the operator size is equal to $2$ for the circuits with two parameterized gates. As we increase the number of parameterized (and entangling gates), the maximum value of the operator size is increased for the 3-qubit, but quickly reaches a plateau. Once again, these results suggest that the 3-qubit circuit structure is more expressive than the 2-qubit.
\subsection{Teacher-student scheme}\label{TS_scheme}

To further analyze the expressive power of the two models in Fig.~\ref{simple_circuits}, we employ the Teacher-student scheme introduced in Section~\ref{TS_theory}. The analysis of the differences $\overline{\Delta y}$ of the prediction maps validates the results from the average operator size in the previous section. 

We compute the $\overline{\Delta y}$ defined in Eq.~\ref{dif} and see that on average both students are approximately $30 \% $ off from the desired target distributions of their teachers. An example of the prediction maps are shown in Fig.~\ref{TS} in the Appendix~\ref{Prediction_maps}, which suggests that the student learns pretty much uncorrelated labeling. These results are in accordance with the results from the averaged operator size of Fig.~\ref{av_results} for two parameterized gates in the sense that both of them have similar values for their averaged operator size, i.e. $2$ and approximately $1.7$ for the 2-qubit and 3-qubit students, respectively.

\begin{figure}[h]
\includegraphics[width=\columnwidth]{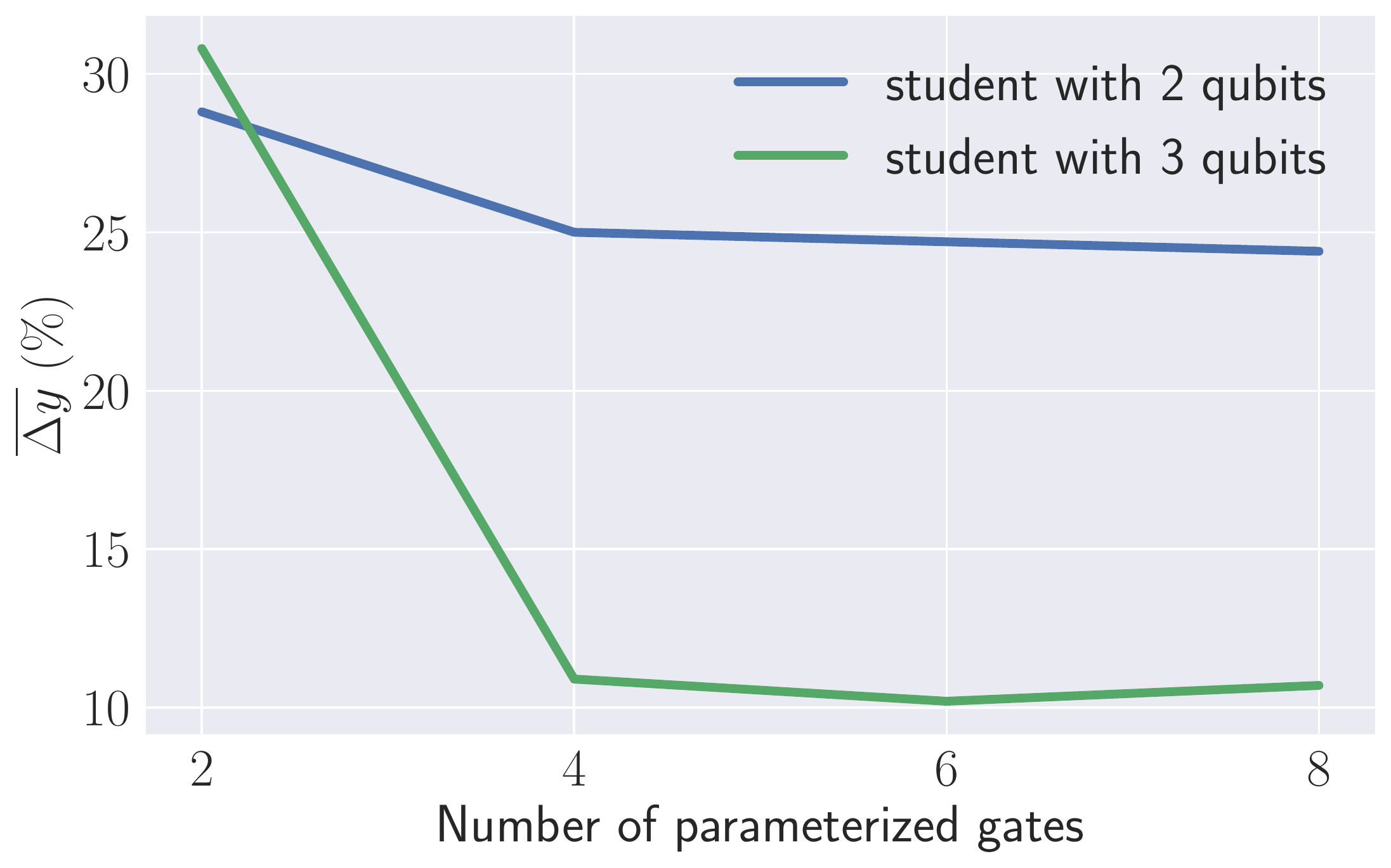}
\caption{We plot the average of the differences between the prediction maps of the teacher and student defined in Eq.\eqref{diffs} as we increase the number of parameterized and entangling gates for the simple circuits shown in Fig.\ref{simple_circuits}. The blue line is for the Teacher with 3 qubits and the student with 2, while the orange line is for the opposite case. }
\label{diffs} 
\end{figure}

Next, we increase the number of parameterized (and entangling gates) by adding a layer $L$ and present the results in Fig.~\ref{diffs}. We see a significant improvement in performance by increasing the parameterized gates from 2 to 4 for the 3-qubit student. This means that it can learn more reliably the labeling provided by the 2-qubit teacher. On the contrary, the 2-qubit student has almost a constant $\overline{\Delta y}$ for any amount of parameterized gates which suggests that it can't improve and learn more reliably the 3-qubit teacher's outputs.

To obtain these results we used a dataset of 500 points on a 2D grid with $x_i \in [-\pi, \pi]$ on a grid and we generated 100 different labelings via different random initalizations of the teacher. The students are as well randomly initiallized and trained until convergence. These results are in agreement with the calculations from the averaged operator size of the previous subsection~\ref{av_analysis}. In both cases, for the 3-qubit circuit, we observe an improvement in performance as we increase the number of parameterized gates. Finally, we quickly reach a plateau in performance as we increase further the number of gates as it is also observed in the analysis of the averaged operator size.
%
\subsection{The type of functions that simple quantum models learn}\label{Fourier section}
The results from the previous two subsections $\{$~\ref{av_analysis}~\ref{TS_scheme}$\}$ suggest that simple changes in the structure of the circuit highly affects the average operator size and $\overline{\Delta y}$, i.e. the learning capability and expressive power of the models. Therefore, the number of parameterized (and entangling) gates, the circuit architecture and the number of (ancilla) qubits play an important role in QML model's performance. To further analyze the effect of the processing and measurement operators on the expressive power of the models, we use the representation of quantum models as partial Fourier series introduced in subsection~\ref{Fourier}. 

The Fourier coefficients defined in Eq.\ref{Ffun} are determined by the eigenvalues of the data-encoding Hamiltonians. For example, the data-encoding Hamiltonian for each qubit from the circuits of Fig.~\ref{simple_circuits} is $H=\dfrac{1}{2} \sigma_x$ and we can assume without loss of generality that it has two distinct eigenvalues, i.e. $\lambda_1 = -1, \lambda_2 = 1$. Importantly, since we have the same type of encoding gate for each qubit, they have the same frequency spectrum $\Omega$. As it is explained in detail in the work of \cite{schuld2020effect}, we can derive the frequency spectrum for each qubit $\Omega=\{-1,0,1\}$ from the possible differences $n_{qubit} = \lambda^{qubit}_{j} - \lambda^{qubit}_{i}$ for $\lambda^{qubit}_{j},\lambda^{qubit}_{i} \in \{-1, 1\}$. Then, we have the Fourier coefficients as $c_{n_1 n_2}$ and Eq.~\eqref{Ffun} can be written as
\begin{equation}\label{our_pFourier}
    f(x)= \sum_{n_1 \in\Omega} \sum_{n_2 \in\Omega} c_{n_1 n_2} e^{-in_1 x_1} e^{-in_2  x_2}. 
\end{equation}
Both circuit models in Fig.~\ref{simple_circuits} have the same number of Fourier basis functions given by Eq.~\ref{our_pFourier}, since they have the same number of encoding gates. This fundamentally limits their learning ability. But importantly, the structure of the processing and measurement operator affect the distribution of the coefficients.

\begin{figure}[h]
\includegraphics[width=\columnwidth]{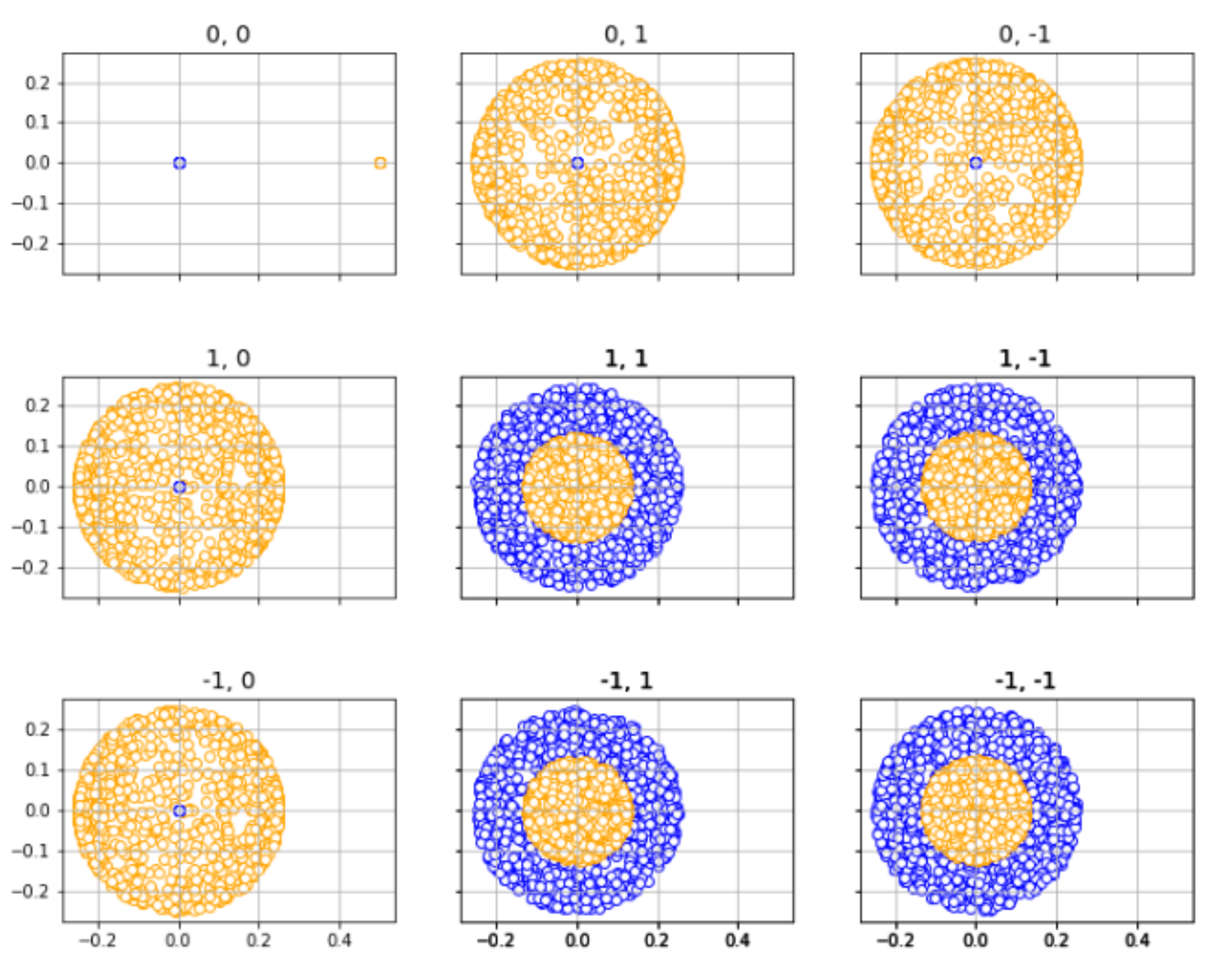}
\caption{The real and imaginary part of the Fourier coefficients for the 2-qubit (blue color) and 3-qubit (orange color) explicit circuits of Fig.~\ref{simple_circuits} with two parameterized gates each. }
\label{coeff_2qubit3} 
\end{figure}

In Fig.~\ref{coeff_2qubit3}, we plot the real and imaginary parts of the Fourier coefficients for the 2-qubit (blue color) and 3-qubit (orange color) explicit circuits of Fig.~\ref{simple_circuits} with two parameterized gates each. 
The coefficients $c_{00}, c_{01}, c_{10}, c_{-10}, c_{0,-1}$ from Fig.~\ref{simple_circuits} for the 2-qubit circuit are all zero (blue color), while they are non-zero for the 3-qubit (orange color). But what happens with the coefficients $c_{11}, c_{-1,-1}, c_{-11}, c_{1,-1}$? Interestingly, computing the exact analytical formula for the distributions of the coefficient $c_{1,1}$ shows that they are different. We present these analytical formulas in Appendix~\ref{coeff}. In summary, even though the two quantum models of Fig.~\ref{simple_circuits} have access to the same Fourier basis functions, the different distribution of their coefficients results in a different combination of these basis functions, and hence, give rise to different function classes.

This is also in accordance with the results from the previous two subsections $\{$~\ref{av_analysis}~\ref{TS_scheme}$\}$. The 2-qubit student (Fig.\ref{simple_circuits} left) has difficulties learning the prediction map of the 3-qubit teacher (Fig.\ref{simple_circuits} right), since it has four coefficients strictly equal to zero which does not help approximating the more complicated distribution of the 3-qubit teacher. But reversing the roles, the 3-qubit student has also difficulties learning the 2-qubit teacher. Even though, the student could learn the zero coefficients, the coefficients $c_{11}, c_{-1,-1}, c_{-11}, c_{1,-1}$ of the teacher belong to a different function class, i.e. they draw their values from a different distribution. This could be seen from the larger spread of the coefficients in Fig.~\ref{coeff_2qubit3}. These results are in accordance with the averaged operator size (Fig.\ref{av_results}) and $\overline{\Delta y}$ (Fig.\ref{diffs}), since they also suggest that for just two parameterized gates both the 2-qubit and 3-qubit circuit have similar performance.

Most importantly, once we add an extra layer $L$ to the 3-qubit circuit, the spread of the Fourier coefficients in Figure~\ref{coeff_2qubit3} (in orange) increases and overlaps with the coefficients of the 2-qubit circuit (in blue) almost completely. These results are shown in light blue color at Fig.\ref{3qubits_4gates} in the Appendix~\ref{extra_plot}). This is in accordance with the averaged operator size and $\overline{\Delta y}$ from the teacher-student scheme where a better performance is observed for the 3-qubit case by adding an extra layer $L$. On the contrary, adding an extra layer in the 2-qubit case does not improve the performance of the student as suggested as well from Fig.\ref{av_results} and Fig.\ref{diffs}, i.e. the distribution of the coefficients shown in Figure~\ref{coeff_2qubit3} in blue stays the same.

But what more can we say for the distribution of these coefficients? Could we understand which circuit elements determine the exact zero terms? We tackle these questions in the next section.
\section{Variational measurement }\label{Two_parts}
\begin{figure}[h]
\includegraphics[width=\columnwidth]{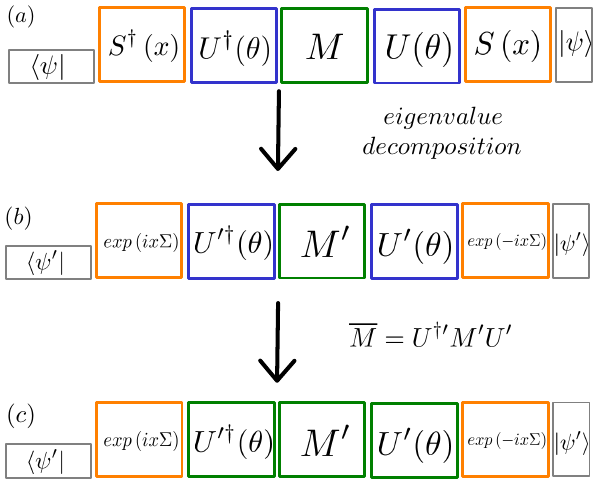}
\caption{We illustrate the matrix block representation of the expectation value of any circuit given by Fig.~\ref{parts} (a), after following the Fourier analysis, i.e. with the eigenvalue decomposition of the encoding unitaries (orange color) (b) and after considering as one the processing and measurement operators shown in green color (c). }
\label{matrix_blocks} 
\end{figure}

We can describe the expectation value of any circuit of Fig.~\ref{parts} with the representation of matrix blocks shown in Fig.\ref{matrix_blocks}a. 
Following~\cite{schuld2020effect}, the data encoding unitary can be written as $S(x)=V^{\dagger} e^{-i x \Sigma} V$ via an eigenvalue decomposition, where $\Sigma$ is a diagonal matrix of eigenvalues and $V$ is the unitary formed with the eigenvectors as columns. We absorb $V, V^{\dagger}$ into the initial state $\ket{\psi '} = V \ket{\psi}$ and into the processing part of the circuit $U^{\prime}=V U V^{\dagger}$. For consistency, we transform the measurement operator with $M^{\prime} = V M V^{\dagger}$. As a result, the encoding unitary is simply a diagonal matrix $\exp(-i x \Sigma)$. The transformed matrix blocks are shown in Fig.~\ref{matrix_blocks}b. 

\begin{figure}[h]
\includegraphics[width=\columnwidth]{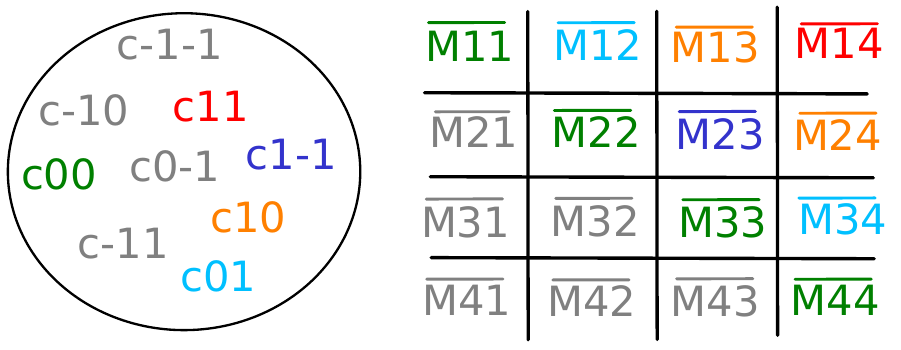}
\caption{The mapping between the Fourier coefficient and the matrix elements of an arbitrary matrix $\overline{M}$. The colors represent the five independent Fourier coefficients from the nine coefficients in total, if we exclude the complex conjugates shown in grey color.}
\label{mapping} 
\end{figure}

In Fig.~\ref{mapping}, we schematically illustrate the mapping between the Fourier coefficients and the frequencies instead of explicitly writing the full matrix given by the inner product of the orange, blue and green matrix blocks of Fig.~\ref{matrix_blocks}b. The Fourier coefficients depend only on the processing and measurement operators, therefore the matrix elements of $ \overline{M} = U^{\dagger '} M' U'$, while the frequencies on the encoding operator, i.e. the diagonal matrix of the encoding Hamiltonian as explained in detail in~\cite{schuld2020effect}. As introduced in Eq.~\ref{our_pFourier}, the Fourier coefficients are defined as $c_{n_1, n_2}$. For example, $c_{00}$ corresponds to the frequencies with $n_1=n_2=0$ for both qubits, which after explicitly calculating the expectation value in Eq.~\ref{our_pFourier} in turn correspond to the diagonal  matrix elements of $\overline{M}$. Following the same procedure, $c_{01}$ is the sum of elements $\overline{M}_{12}$ and $\overline{M}_{34}$, $c_{10}$ is the sum of elements $\overline{M}_{13}$ and $\overline{M}_{24}$, $c_{11}$ is the element $\overline{M}_{14}$ and $c_{1,-1}$ is the element $M_{23}$. Finally, $c_{0,-1}$, $c_{-10}$, $c_{-1-1}$ and $c_{-11}$ are the complex conjugates of $c_{01}$,  $c_{10}$, $c_{11}$ and $c_{1-1}$, respectively. 

We now discuss the simple example of the 2-qubit case from Fig.~\ref{simple_circuits} with the Pauli Z operator applied to the 2nd qubit ($M'_1 = I$, $M'_2 = Z$). The transformed measurement operator $M' = M^{\prime}_1 \otimes M^{\prime}_2 = I \otimes \left( VZV^{\dagger} \right)$ of the data-encoding Hamiltonian) becomes
\begin{align}
\begin{matrix}\label{M_meas}
    M' = \left[\begin{array}{llll}
0 & h & 0 & 0 \\
k & 0 & 0 & 0 \\
0 & 0 & 0 & h^{*} \\
0 & 0 & k^{*} & 0
\end{array}\right],
\end{matrix}
\end{align}
where $h, k$ are complex trigonometric functions that determine the distribution of the coefficients. But the mapping between the zero elements and the coefficients that are zero does not match. For example, $c_{11}$ which is determined by $\overline{M}_{14}$ has a non zero distribution as shown in Fig.\ref{coeff_2qubit3} (blue color), while eq.~\eqref{M_meas} suggests that $\overline{M}_{14}$ it should be strictly zero.

To resolve this discrepancy, we need to consider the processing and measurement operators as one matrix block $\overline{M} = U^{\dagger \prime} M^{\prime} U^{\prime} $ (shown in green in Fig.~\ref{matrix_blocks}c). This is in accordance with recent works~\cite{schuld2021quantum,  Variational_meas}, where the processing and measurement operators are regarded as one part in the circuit structure.

For the 2-qubit case, the combined matrix $\overline{M}$ (green blocks in Fig.\ref{matrix_blocks}c) becomes:
\begin{align}
\begin{matrix}\label{M_cnot}
    \overline{M} =\left[\begin{array}{llll}
0 & 0 & 0 & f \\
0 & 0 & g & 0 \\
0 & g* & 0 & 0 \\
f* & 0 & 0 & 0
\end{array}\right],
\end{matrix}
\end{align}
where $f, g$ are complex trigonometric functions that determine the distribution of the coefficients shown in Fig. \ref{coeff_2qubit3}. Following the mapping between the coefficients and the elements of the combined matrix $\overline{M}$, we can immediately see which coefficients are zero. The only non-zero coefficients are the $c_{11}$, $c_{1,-1}$ and their complex conjugates determined by $\overline{M}_{14}$, $\overline{M}_{23}$ and $\overline{M}_{41}$,$\overline{M}_{32}$, respectively. 

This analysis suggests that in some cases replacing the usual segregation of a quantum circuit into three parts: encoding, processing and measurement with just two parts: encoding and variational measurement could reveal further insights on the performance and properties of QML models. 

\subsection{Toy application}
\begin{figure}[h]
\includegraphics[width=\columnwidth]{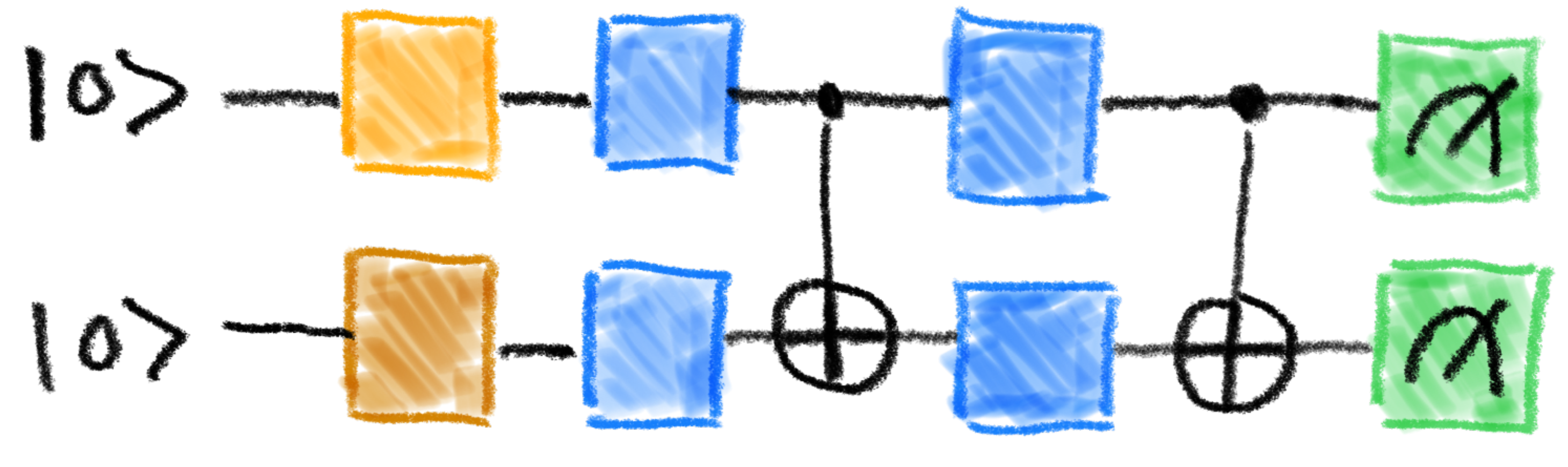}
\caption{A 2-qubit circuit with 4 parameterized and two entangling gates where we could either measure on the 1st or 2nd qubit.}
\label{2qubits_4gates} 
\end{figure}

\begin{figure}[h]
\includegraphics[width=\columnwidth]{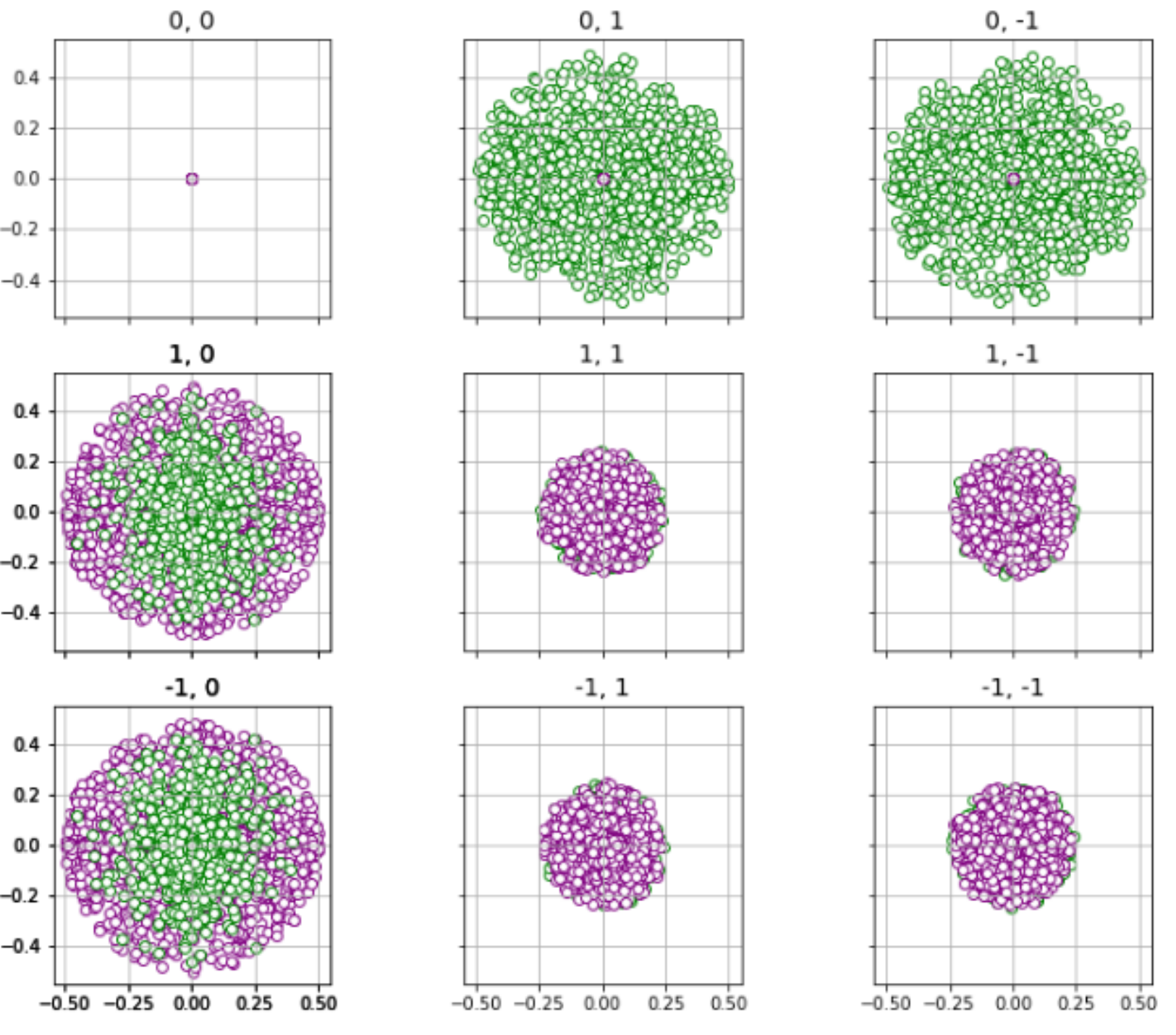}
\caption{The real and imaginary part of the Fourier coefficients of the circuit structure shown in  Fig.~\ref{2qubits_4gates}. If the circuit is measure on the first qubit we obtain the Fourier coefficients in purple, if we measure on the second qubit we obtain the coefficients in green. }
\label{meas} 
\end{figure}

To further emphasize the importance of the measurement operators of a circuit, we present a simple quantum model with two qubits (Fig.\ref{2qubits_4gates}). The Fourier coefficients of this circuit are represented in Fig.\ref{meas}, where the coefficients in purple represent the measurement of the 1st qubit and the Fourier coefficients in green represent the measurement on the 2nd qubit. 
As we see, the distribution of the Fourier coefficients differ greatly. Therefore, with the same circuit structure, we could create two different functions classes depending on where we measure. This might have a more relevant applications, but it already suggests that once we better understand how simple elements on our circuit structure affect the overall performance it is more natural to think of applications.
\section{Conclusions}
\label{conclusions}
The data encoding is fundamentally limiting the performance of a circuit, but the processing and measurement can also significantly affect the capabilities of the model. We show that different parts of processing and measurement operators affect the performance of the QML model. As expected, increasing the number of layers $L$ has a direct effect on the model's performance, but a plateau in the performance seems to be reached. We also observed that the circuit with the ancilla has overall a better performance in accordance with the recent work of~\cite{ancilla_best}. 

To quantify the model's performance, we introduced three different tools: the averaged operator size, the $\overline{\Delta_y}$ and the coefficient from the partial Fourier series. By applying the Fourier representation, we see that two different circuits with the same number of parameters could give rise to different function classes. Finally, we find that the segregation of the processing and measurement operator might not be ideal when interpreting the parts of the circuit. Instead when considered as one, i.e. the variational measurement, could give insights on the model's performance. The idea of the variational measurement has also been presented in the context of quantum kernels \cite{Schuld_kernel}. This might bring the relation between the Fourier representation and quantum kernels one step closer. 

Once we better understand how QML models work, we can naturally start thinking about applications. For example, we presented a trivial implementation that arises from where we measure. Specifically, measuring either one of the two qubits on a 2-qubit circuit gives rise to fundamentally different classes of functions that this model could express. Therefore, with the same circuit structure one could learn or express two different function classes. This can be seen from the numerical calculation of the real and imaginary part of the Fourier coefficients in Fig.~\ref{2qubits_4gates}. 

We understand that computing the analytical trigonometric expression for more qubits, or even gates, is similar to opening Pandora's box. Instead we suggest to focus on the other two tools proposed, i.e. the averaged operator size and the map differencies from the teacher-student scheme. In this work, we focused on simple quantum models and verified that the tools presented here are robust. The analytical and numerical results are in agreement, and therefore, these tools could be further used to test the performance of more complex circuit structures.

In conclusion, applying the averaged operator size and teacher-student scheme to other simple circuit structures could further help understand how QML models work. It would also be interesting to exploit the effect of measurement for a more interesting use-case application. But to do so, a careful study on how the measurement position affects the model's performance should be undertaken, i.e. by increasing the number of qubits and circuit complexity~\cite{Haferkamp_2022}. This study could also further strengthen the belief that ancillary qubits have better performance overall \cite{ancilla_best}.

Now, we have a better understanding on the basic questions posed earlier in the introduction: "How many parameterized gates to use?", "How many entangling gates and where?", "Do ancilla qubits help?"  and "Where to measure?". We understand that the processing and measurement operators strongly affect the performance of the QML model. Therefore, focusing the research directions on how exactly they affect it is of great importance. If we want to find the great applications that QML promises~\cite{q_advantage, quantum_advantage}, we first need to understand how QML models work.

\section{Code}
Code to reproduce the results and explore further settings can be found in the following Github repository: 
\url{https://github.com/KaterinaGratsea/Teacher-student_scheme-part-2}.

\section{Acknowledgements}
The simulations were made with the Pennylane library~\cite{bergholm2018pennylane} and the graphics with the \kat{?} software.
ICFO group acknowledges support from: ERC AdG NOQIA; Agencia Estatal de Investigación (R\&D project CEX2019-000910-S, funded by MCIN/ AEI/10.13039/501100011033, Plan National FIDEUA PID2019-106901GB-I00, FPI, QUANTERA MAQS PCI2019-111828-2, QUANTERA DYNAMITE PCI2022-132919,  Proyectos de I+D+I “Retos Colaboración” QUSPIN RTC2019-007196-7), MCIN via European Union NextGenerationEU (PRTR);  Fundació Cellex; Fundació Mir-Puig; Generalitat de Catalunya through the European Social Fund FEDER and CERCA program (AGAUR Grant No. 2017 SGR 134, QuantumCAT \ U16-011424, co-funded by ERDF Operational Program of Catalonia 2014-2020); EU Horizon 2020 FET-OPEN OPTOlogic (Grant No 899794); National Science Centre, Poland (Symfonia Grant No. 2016/20/W/ST4/00314); European Union’s Horizon 2020 research and innovation programme under the Marie Skłodowska-Curie grant agreement No 847517.

\renewcommand{\bibsection}{\section*{References}}

\bibliography{main.bib}

\appendix 

\section{Monte-Carlo integration}\label{MC}
For the numerical simulation we use Monte-Carlo integration~\cite{McClean} and approximate the integral as 
\begin{equation}
\int_{U(N)} f(U) d U \approx \frac{1}{p} \sum_{i=1}^{p} f\left(U_{i}\right),
\end{equation}
where $p$ is the total number of random unitaries used and $U_i$ is a randomly drawn unitary according to the Haar measure. To define such unitary numerically, we start by calling a $N \times N$ matrix with Gaussian values. Then, we perform a QR decomposition on this matrix which gives two matrices $Q$ and $R$. Next, we define the diagonal matrix D from the diagonal elements of the matrix $R$, i.e. $D_{i i}=R_{i i} /\left|R_{i i}\right|$. Finally, the unitary random matrix according to the Haar measure is defined as $U_i = QD$~\cite{McClean}.

\section{Prediction maps}\label{Prediction_maps}
\begin{figure}[h]
\includegraphics[width=\columnwidth]{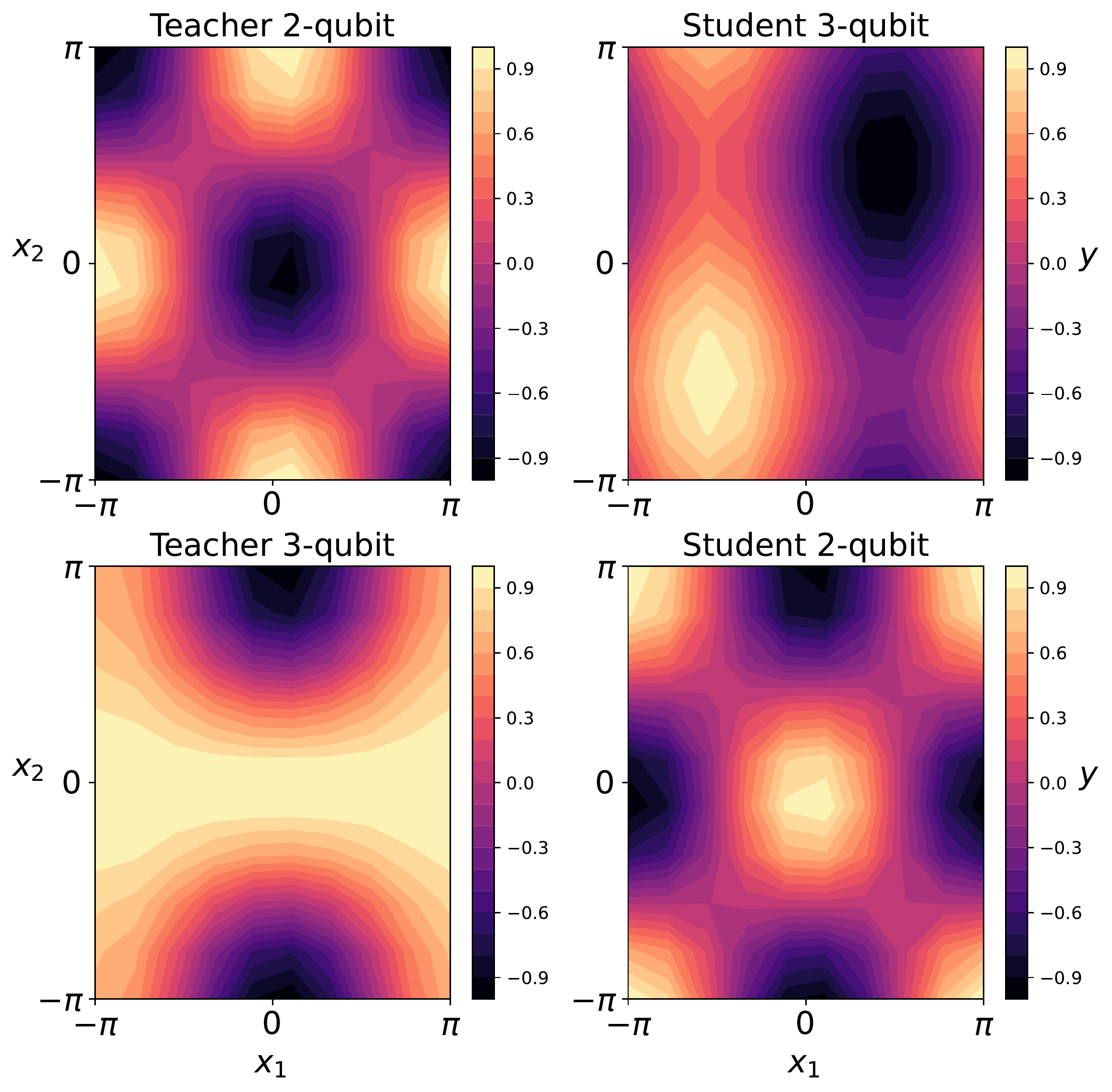}
\caption{Examples of the prediction maps for the 2-qubit and 3-qubit circuits at the roles of the teacher and student. In both cases, the students have significant discrepancies with their teachers.}
\label{TS} 
\end{figure}

Fig.~\ref{TS} shows an example of the prediction maps for the 2-qubit circuit as a teacher and the 3-qubit circuit as a student along with the reverse roles for each circuit. In both cases, the student deviates from the data distribution of its teacher. 

\section{Coefficient $c_{11}$}\label{coeff}

Here, we present the explicit formula of the real and imaginary parts of the coefficient $c_{11}$ for the 2-qubit and 3-qubit circuits in Fig.~\ref{simple_circuits} in Eq.~\eqref{Re_c11},\eqref{Im_c11} and Eq.~\eqref{Re3_c11},\eqref{Im3_c11}, respectively for the real and imaginary parts.
\begin{widetext}
\begin{multline}\label{Re_c11}
\begin{array}{l}
 Re\left( c_{11} \right) =\sin \left(\frac{\theta_2}{2}\right) \sin \left(\frac{\phi_1}{2}+\frac{\omega_1}{2}\right) \cos \left(\frac{\theta_1}{2}\right) \cos \left(\frac{\theta_2}{2}\right) \sin \left(\frac{\theta_1}{2}\right) \cos \left(\frac{-\phi_1}{2}+\frac{\omega_1}{2}\right) \\ \left(\sin \left(\frac{\phi_2}{2}+\frac{\omega_2}{2}\right) \cos \left(-\frac{\phi_2}{2}+\frac{\omega_2}{2}\right) 
 -\cos \left(\frac{\phi_2}{2}+\frac{\omega_2}{2}\right) \sin \left(-\frac{\phi_2}{2}+\frac{\omega_2}{2}\right)\right)  \\
 -\sin \left(\frac{\theta_2}{2}\right) \sin \left(-\frac{\phi_1}{2}+\frac{\omega_1}{2}\right) \cos \left(\frac{\theta_1}{2}\right) \cos \left(\frac{\theta_2}{2}\right) \sin \left(\frac{\theta_1}{2}\right) 
  \cos \left(\frac{\phi_1}{2}+\frac{\omega_1}{2}\right) \\
 \left(\sin \left( \frac { \phi_2 } { 2 } + \frac { \omega_2 } { 2 } \right) \cos \left(-\frac{\phi_2}{2}+\frac{\omega_2}{2}\right)-\cos \left(\frac{\phi_2}{2}+\frac{\omega_2}{2}\right) \sin \left(-\frac{\phi_2}{2}+\frac{\omega_2}{2}\right)\right)+\frac{1}{4}\left(-4 \cos \left(\frac{\theta_2}{2}\right)^{2}+2\right) \cos \left(\frac{\theta_1}{2}\right)^{2}+ \frac{1}{2} \cos \left(\frac{\theta_2}{2}\right)^{2}-\frac{1}{4}
\end{array}
\end{multline}
\end{widetext}
\begin{widetext}
\begin{multline}\label{Im_c11}
\begin{array}{l}
 Im\left( c_{11} \right)= -\left(\cos \left(\frac{\theta_2}{2}\right)^{2}-\frac{1}{2}\right) \sin \left(\frac{\theta_1}{2}\right) \sin \left(\frac{\phi_1}{2}+\frac{\omega 1}{2}\right) \cos \left(\frac{\theta_1}{2}\right) \cos \left(-\frac{\phi_1}{2}+\frac{\omega 1}{2}\right) \\
 +\left(\cos \left(\frac{\theta_2}{2}\right)^{2}-\frac{1}{2}\right) \sin \left(-\frac{\phi_1}{2}+\frac{\omega 1}{2}\right) \sin \left(\frac{\theta_1}{2}\right) \cos \left(\frac{\theta_1}{2}\right) \cos \left(\frac{\phi_1}{2}+\frac{\omega 1}{2}\right)-\sin \left(\frac{\theta_2}{2}\right) \cos \left(\frac{\theta_2}{2}\right) \\  \left(\sin \left(\frac{\phi_2}{2}+\frac{\omega 2}{2}\right) \cos \left(-\frac{\phi_2}{2}+\frac{\omega 2}{2}\right) 
 -\cos \left(\frac{\phi_2}{2}+\frac{\omega 2}{2}\right) \sin \left(-\frac{\phi_2}{2}+\frac{\omega 2}{2}\right)\right)\left(\cos \left(\frac{\theta_1}{2}\right)^{2}-\frac{1}{2}\right)
\end{array}
\end{multline}
\end{widetext}

\begin{widetext}
\begin{multline}\label{Re3_c11}
\begin{array}{l}
 Re\left( c_{11} \right)
= -\frac{1}{8}-\frac{1}{8}\left(-2+4 \cos \left(\frac{\theta_2}{2}\right)^{2}\right) \cos \left(\frac{\theta_1}{2}\right)^{2}+\frac{1}{2}\sin (\phi_2) \sin (\theta_1) \sin \left(\frac{\theta_1}{2}\right) \sin \left(\frac{\theta_2}{2}\right) \cos \left(\frac{\theta_2}{2}\right) \cos \left(\frac{\theta_1}{2}\right)+\frac{1}{4}\cos \left(\frac{\theta_2}{2}\right)^{2}
\end{array}
\end{multline}
\end{widetext}

\begin{widetext}
\begin{multline}\label{Im3_c11}
\begin{array}{l}
 Im\left( c_{11} \right)=  \frac{1}{2}\sin (\phi_2) \sin \left(\frac{\theta_2}{2}\right) \cos \left(\frac{\theta_2}{2}\right) \cos \left(\frac{\theta_1}{2}\right)^{2}-\frac{1}{2}\sin (\phi_1)\left(\cos \left(\frac{\theta_2}{2}\right)^{2}-\frac{1}{2}\right) \sin \left(\frac{\theta_1}{2}\right) \cos \left(\frac{\theta_1}{2}\right)+\frac{1}{4}\sin \left(\frac{\theta_2}{2}\right) \cos \left(\frac{\theta_2}{2}\right) \sin (\theta_2)
\end{array}
\end{multline}
\end{widetext}

\section{3-qubit circuit with 4 gates}\label{extra_plot}
\begin{figure}[h]
\includegraphics[width=\columnwidth]{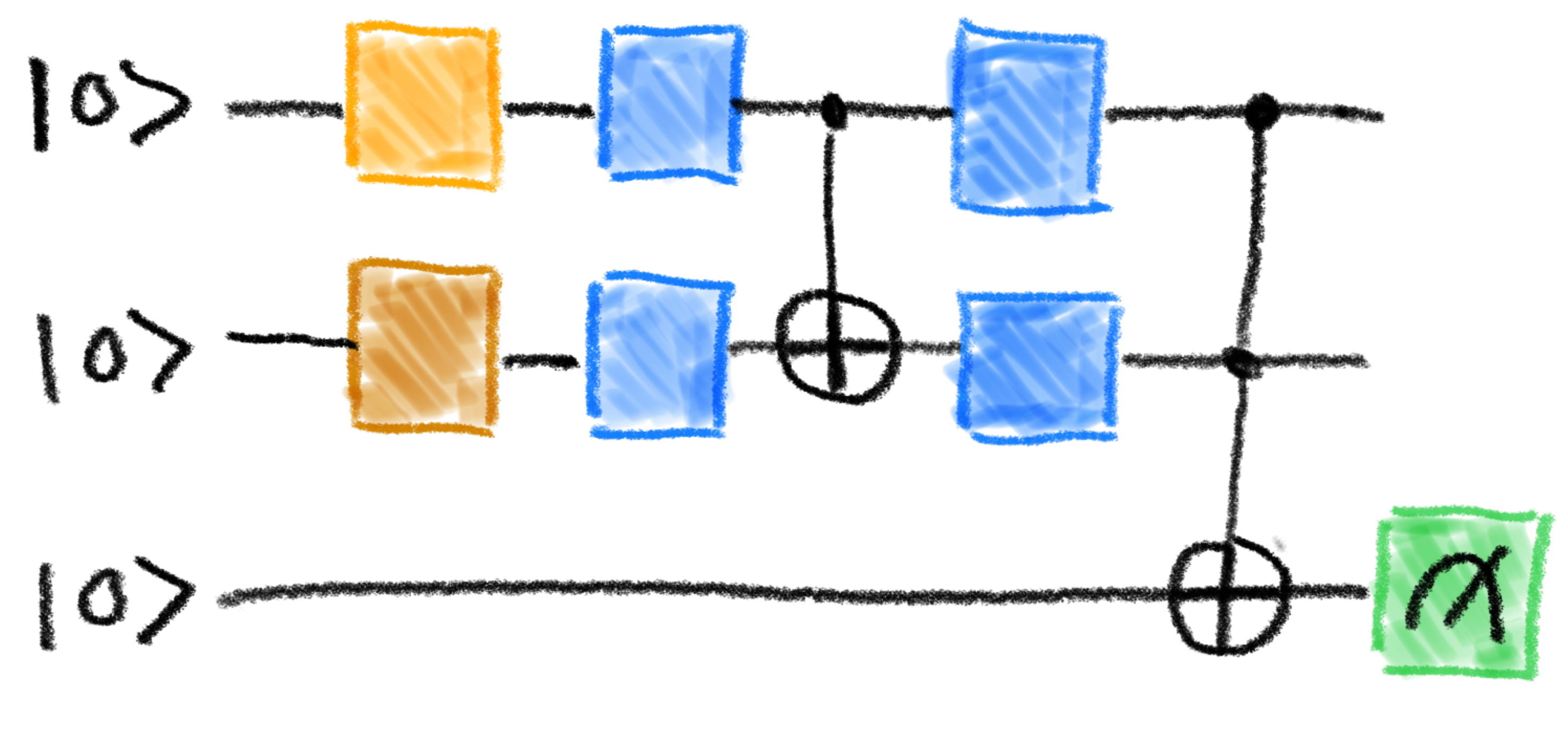}
\caption{A 3-qubit circuit with an extra layer $L$ marked by the dashed lines.}
\label{3qubits_4gates_circ} 
\end{figure}

\begin{figure}[h]
\includegraphics[width=\columnwidth]{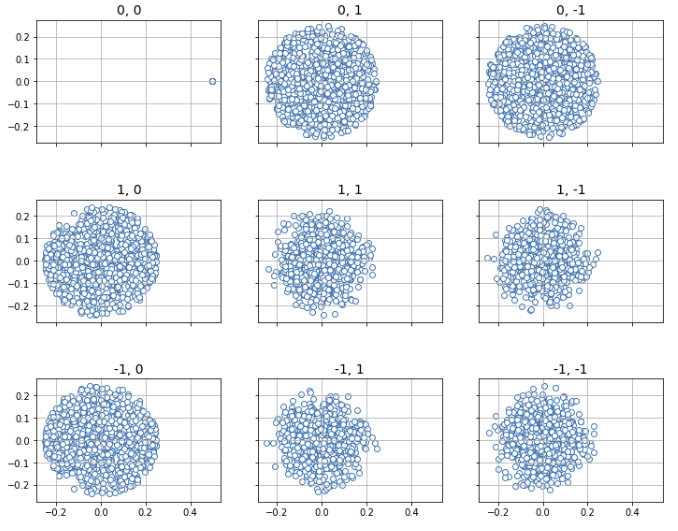}
\caption{The real and imaginary part of the Fourier coefficients for the 3-qubit (light blue color) circuit of Fig.~\ref{3qubits_4gates_circ}. }
\label{3qubits_4gates} 
\end{figure}

Here, we present the Fourier coefficients in Fig.~\ref{3qubits_4gates} for the circuit with 3 qubits and four parameterized gates sketched in Fig.~\ref{3qubits_4gates_circ}. We see that the coefficients $c_{1,1}$ and $c_{1,-1}$ now take similar values with their corresponding coefficients shown in Fig.~\ref{coeff_2qubit3} in orange color.
\end{document}